\documentclass[pss_cm,fleqn]{w-art_cm}
\usepackage{times,w-thm,graphicx,bm,amsmath,amssymb}

\newcommand{\Eq}[1]{Eq.~\eqref{#1}}
\newcommand{\pdag}{{\phantom{\dagger}}}

\newcommand{\past}{{\phantom{\ast}}}

\newcommand{\beq}{\begin{equation}}
\newcommand{\eeq}{\end{equation}}

\newcommand{\beqa}{\begin{eqnarray}}
\newcommand{\eeqa}{\end{eqnarray}}

\newcommand{\PRB}[1]{Phys. Rev. B~\textbf{#1}}
\newcommand{\PRL}[1]{Phys. Rev. Lett.~\textbf{#1}}
\newcommand{\RMP}[1]{Rev. Mod. Phys.~\textbf{#1}}
\newcommand{\PR}[1]{Phys. Rev.~\textbf{#1}}
\newcommand{\Nature}[1]{Nature~\textbf{#1}}
\newcommand{\Science}[1]{Science~\textbf{#1}}
\newcommand{\ZhETF}[1]{Zh. Eksp. Teor. Fiz.~\textbf{#1}}
\newcommand{\JETP}[1]{Sov. Phys. - JETP~\textbf{#1}}
\newcommand{\JPCM}[1]{Journ. of Physics: Condens. Matter~\textbf{#1}}
\newcommand{\etal}{et al.} 

\begin{document}
\renewcommand{\copyrightyear}{2005}

\title[Kondo effect in nanostructures]{Kondo Effect in Nanostructures}

\author[M. Pustilnik]{Michael Pustilnik} 
\address{School of Physics, Georgia Institute of Technology,
Atlanta, GA 30332, USA}



\begin{abstract}
Kondo effect arises whenever a coupling to the Fermi gas induces
transitions within the otherwise degenerate ground state multiplet of an
interacting system. Both coupling to the Fermi gas and interactions are
naturally present in any nanoscale transport experiment. At the same
time, many nanostructures can easily be tuned to the vicinity of a
degeneracy point. This is why the Kondo effect in its various forms 
often influences the low temperature transport in meso- and nanoscale
systems. In this short review we discuss the basic physics of the Kondo 
effect and its manifestations in the low-temperature electronic transport through 
a single electron transistor.
\end{abstract}

\maketitle 

In a typical transport experiment a nanostructure is connected via 
tunneling junctions to two massive conducting electrodes. The 
differential conductance $dI/dV$ of such device often exhibits 
an enhancement when the temperature $T$, magnetic field $B=g\mu_B H$, 
and the bias $V$ are lowered. This enhancement is usually very well 
described by a simple formula 
\beq
dI/dV \propto \frac{\,e^2}{h}
\left[\ln\frac{\max\lbrace T,eV,B\rbrace}{T_K}\right]^{-2},
\label{1}
\eeq
where $T_K$ is the characteristic energy scale of the effect. Such
anomalous behavior was observed in a variety of systems. The list (by no means 
exhaustive) includes lateral semiconductor quantum 
dots~\cite{lateral,lateral_unitary,two-stage}, vertical quantum 
dots~\cite{vertical}, carbon nanotubes~\cite{nanotube,nanotube-orbital}, 
single-molecule transistors~\cite{molecule}, STM imaging of single 
magnetic atoms adsorbed on metallic surfaces~\cite{adatom}, ultra-small
clusters formed in metallic break junctions~\cite{break-junction}, as well 
as more exotic objects such as an antidot~\cite{antidot} (a compressible 
region formed about a potential hill in a two-dimensional electron gas in 
the quantum Hall regime).   
The logarithmic increase of the transport coefficient, such as the differential 
conductance in  \Eq{1}, is nothing new in condensed matter physics. In fact, 
it has the same origin~\cite{review} as the well-known non-monotonic 
temperature dependence of the resistivity of a metal containing magnetic 
impurities -- the \textit{ Kondo effect}. 

\section{Conventional Kondo effect}

The Kondo effect represents a rare (if not unique) example of the 
phenomenon named after a theorist who was the first to explain it~\cite{Kondo}.
The \textit{resistance minimum}, as the Kondo effect was known prior to 1964, 
was discovered in the early 1930s~\cite{deHaas}. Later on (historical 
account can be found in~\cite{Kondo_05,Anderson_history,Hewson}), 
the impurity contribution to the resistivity of dilute alloys was cast in 
the form of an empirical law
\beq
\delta\rho\,(T) \propto n_i \ln(\epsilon_F\!/T),
\label{2}
\eeq
where $n_i$ is the impurity concentration. Since the contribution to 
the resistivity from the electron-phonon scattering decreases with the 
decrease of $T$, \Eq{2} leads to a minimum of the resistivity at a 
certain temperature. By 1964, there has been accumulated a considerable 
amount of experimental data suggesting that the effect appears only when 
the impurity atoms are \textit{magnetic}~\cite{Kondo_05}. 
Moreover, the relation $\delta\rho\propto n_i$ has been verified down 
to the lowest attainable impurity concentrations, thereby establishing that 
the phenomenon is a single impurity effect rather than arising from the 
interaction between the impurities. These observations~\cite{Kondo_05} 
led Kondo to consider the simplest possible model describing the exchange 
interaction of a magnetic impurity with the conduction electrons in the host 
metal,
\beq
H_K = H_0 + J\!\left({\bf s}\cdot\!{\bf S}\right),  
\label{3}
\eeq
where $H_0 = \sum_{ks}\xi^\pdag_k \psi^\dagger_{ks} \psi^\pdag_{ks}$
describes the electron gas, ${\bf s} = \frac{1}{2}\sum_{kk'ss'} 
\psi^\dagger_{ks} \hat{\bm{\sigma}}_{ss'}\psi^\pdag_{k's'}$ is the spin 
density of the conduction electrons at the impurity site 
(with $\hat{\bm{\sigma}} = (\hat\sigma^x,\hat\sigma^y,\hat\sigma^z)$ 
being the Pauli matrices), and $\bf S$ is a spin-$1/2$ operator representing 
the magnetic impurity. The phenomenological Kondo model \eqref{3} 
(a.k.a. the $s$-$d$ model) was introduced in the literature at least as early 
as in 1946~\cite{s-d}; its validity was later established~\cite{SW} by 
deriving it from the microscopic Anderson impurity model~\cite{AM}.

Kondo realized~\cite{Kondo_05} that due to the non-commutativity 
of the spin operators in \Eq{3}, treating the exchange in the lowest 
order of perturbation theory is insufficient. Indeed, going beyond the 
Born approximation yields the desired [see \Eq{2}] logarithmic temperature 
dependence already in the third order in the exchange amplitude~\cite{Kondo},
\[
\delta\rho\sim n_i \,(\nu J)^2 \bigl[1+ 2\nu J \ln(D/T)\bigr].
\]
Here $\nu$ is density of states (so that $\nu J\ll 1$ is a dimensionless parameter) 
and $D\sim\epsilon_F$ is the high-energy cutoff in \Eq{3}. This discovery 
solved one of the longest standing puzzles in the history of condensed matter 
physics~\cite{puzzle}. 

Soon after the Kondo's paper \cite{Kondo} was published, it was found that 
logarithmically-divergent contributions appear in all orders of 
perturbation theory, forming a geometric series~\cite{Abrikosov-Suhl} 
\[
\delta\rho(T)/\delta\rho(0)\propto 
\left\lbrace
\sum_{n=0}^\infty(\nu J)^n\bigl[\ln(D/T)\bigr]^{n-1}
\right\rbrace^2
= \left[\frac{\nu J}{1-\nu J\ln(D/T)}\right]^2.
\]
This result can be also written as
\beq
\delta\rho(T)/\delta\rho(0)\propto\bigl[\ln(T/T_K) \bigr]^{-2},
\label{4}
\eeq
cf. \Eq{1}, where the Kondo temperature $T_K$ is given by
\beq
T_K=D\,e^{-1/(\nu J)}.
\label{5}
\eeq
Obviously, \Eq{4} diverges when $T$ approaches $T_K$. Similar untractable 
(and clearly unphysical) divergencies appear in thermodynamic 
quantities as well, indicating the failure of the perturbation theory. 
The problem of dealing with these divergencies became known 
as the \textit{Kondo problem}, and its resolution came later with the 
advent of the powerful Renormalization Group ideas~\cite{PWA,Wilson}.

To get a feeling about the physics of the Kondo model, let us consider 
a cartoon version of it in which the electron gas in \Eq{3} is replaced 
by a single spin-1/2 operator ${\bf S}'$. The ground state of the resulting 
Hamiltonian of two spins $H' = J({\bf S}'\cdot{\bf S})$ is obviously a 
singlet for antiferromagnetic exchange $(J>0)$. The excited state (a triplet) 
is separated from the ground state by the energy gap $J$. This separation 
can be interpreted as the binding energy of the singlet. Unlike ${\bf S}'$ in 
this cartoon example, ${\bf s}$ in \Eq{3} is merely a local spin density of the 
conduction electrons. Because electrons are freely moving in space, it is 
hard for the impurity to ``capture" an electron and form a singlet. Yet, even 
an arbitrarily weak local antiferromagnetic exchange interaction suffices to 
form a singlet ground state~\cite{PWA,Wilson}. However, the characteristic 
energy (an analog of the binding energy) for this singlet is given not by the 
exchange amplitude $J$, but by the Kondo temperature $T_K$, see \Eq{5}. 

This lifting of the degeneracy of the ground state is the very essence 
of the Kondo effect. It is also the origin of the logarithmic divergences 
in perturbation theory. Indeed, in perturbation theory one starts 
with the impurity decoupled from the electron gas $(J\to 0)$. Since the 
spin-up and spin-down states of the impurity are degenerate, the ground state 
of the system in this limit is a doublet. Treating the strength of the exchange 
perturbatively is then justified only at temperatures that significantly 
exceed the singlet binding energy, i.e. at $T\gg T_K$. In the opposite
limit $T\ll T_K$ a version of the perturbation theory that explicitly takes into 
account the correct symmetry of the ground state right from the start yields~\cite{Nozieres} 
\beq
1 - \delta\rho(T)/\delta\rho(0) \propto (T/T_K)^2,
\quad
T\ll T_K.
\label{6}
\eeq
Eqs. \eqref{4} and \eqref{6} are applicable, respectively, in the weak 
$(T\!\gg T_K)$ and strong $(T\!\ll T_K)$ coupling limits. Since the 
Kondo effect is a crossover phenomenon, rather than a phase 
transition~\cite{PWA,Wilson,Bethe}, the function $\delta\rho(T)/\delta\rho(0)$ 
varies smoothly with $T$ in the crossover region $T\!\sim T_K$.

It should be noted that the Kondo effect does not always manifest itself in 
the increase of the resistance. Indeed, the formation of the singlet ground state 
leads to an increase of the probability for an electron to scatter by the impurity.
The closer the energy of the scattered electron is to the Fermi level, the higher 
is the scattering probability.
If a magnetic impurity is imbedded in a bulk sample, the higher scattering 
probability translates to the increase of the resistivity. However, if the impurity 
resides in a tunneling barrier separating two massive conductors, the increase 
of the scattering probability leads to an enhanced probability for an electron to
tunnel through the barrier, hence it is the differential conductance, rather than 
the resistance, that is enhanced with lowering the temperature or bias, cf. \Eq{1}.
Such zero-bias anomalies in tunneling conductance, first observed in the mid 
1960s~\cite{tunneling}, were adequately explained~\cite{Appelbaum}
in the context of the Kondo effect~\cite{tunneling_reviews}.

\section{Coulomb blockade}

By now, the Kondo effect is among the best understood phenomena 
in condensed matter physics. Nevertheless, the interest in the Kondo
problem never really subsided, as it provides an ideal testing ground for
new analytical and numerical methods developed in order to understand 
electronic properties of the materials with strong correlations between the 
electrons. In recent years, this interest received a major boost~\cite{Kondo_popular} 
due to unprecedented advances in experimental techniques associated 
with the advent of nanotechnology. It is now possible to create \textit{artificial}
nanoscale magnetic impurities. Modern experimental methods not only 
offer a direct access to transport properties of such artificial impurities, 
but also provide one with a broad arsenal of tools to tweak the impurity 
properties, unmatched in conventional systems. 

\begin{vchfigure}[htb]
\includegraphics[width=0.38\textwidth]{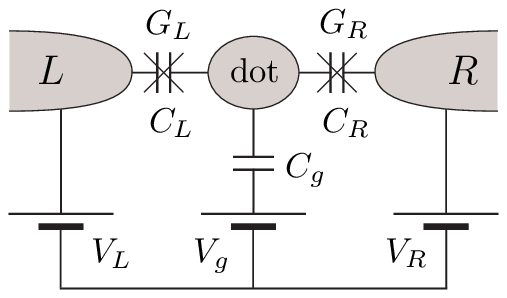}
\caption{
Equivalent circuit for a quantum dot connected to two conducting
leads by tunnel junctions and capacitively coupled to the gate electrode.
}
\label{fig1}
\end{vchfigure}

At the heart of these applications lies the phenomenon of the 
\textit{Coulomb blockade}~\cite{review,blockade_exp,ABG}. 
The foundations of the Coulomb blockade physics were laid in the pioneering 
experimental work~\cite{Giaever} (see also~\cite{LJ,Gorter}). In a typical
\textit{single electron transistor} setup~\cite{blockade_exp} a small system
(say, a quantum dot) is connected via tunneling junctions to two conducting 
electrodes (leads) $R$ and $L$ and is capacitively coupled to the third electrode, 
the gate, see the equivalent circuit in Fig. \ref{fig1}. 

Suppose the dot has charge $q$. The classical electrostatic energy 
associated with this charge is given by
\beq
E(q) = \frac{q^2}{2C} - q \frac{C_g}{C}V_g,
\label{7}
\eeq
where $C=C_L+C_R+C_g$ is the total capacitance of the dot. The 
simplest model Hamiltonian accounting for this energy then reads
\beq
H_\text{dot} = \sum_{ns} \epsilon_n d^\dagger_{ns} d^\pdag_{ns} 
+ E_C\bigl(\hat N - N_0\bigr)^2,
\label{8}
\eeq
where the first term represents the single-particle (noninteracting) part, 
and the second term comes about from \Eq{7} after the replacement
$q\to e\hat N$, where $\hat N = \sum_{ns}d^\dagger_{ns} d^\pdag_{ns}$ 
is the operator of the total number of electrons in the dot. In \Eq{8} 
$E_C=e^2\!/2C$ is the charging energy and $N_0 = C_g V_g/e$ is 
the dimensionless gate voltage. The mean spacing $\delta E$ between 
the single-particle energy levels $\epsilon_n$ is usually small compared 
to $E_C$~\cite{scales}.

\begin{vchfigure}[htb]
\includegraphics[width=0.65\textwidth]{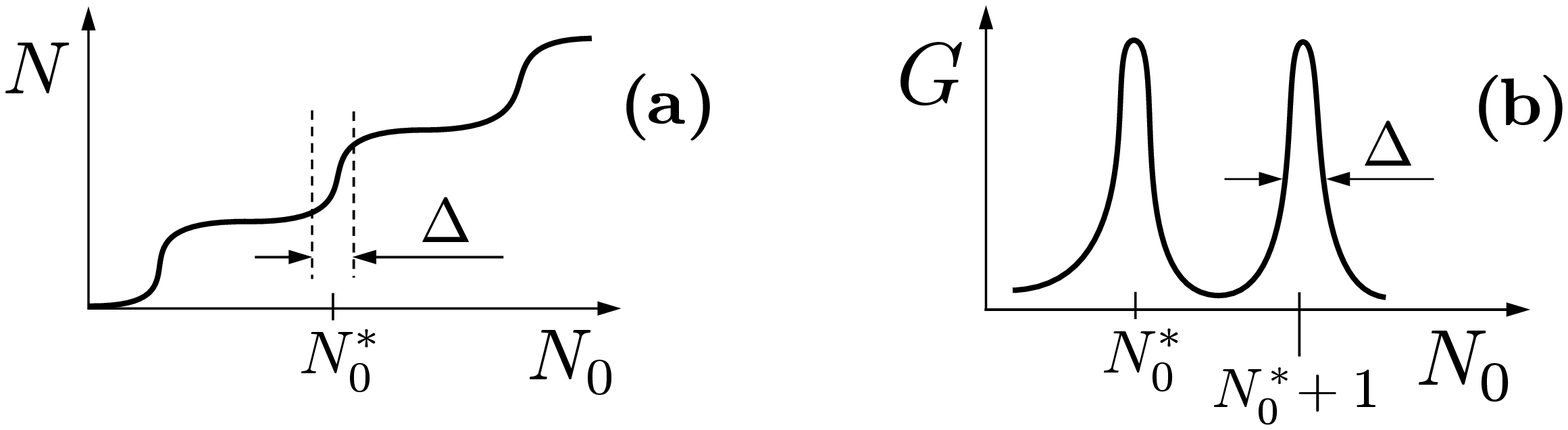}
\caption{
(a) Occupation of the dot $N=\langle\hat N\rangle$ at $T\ll E_C$ 
as function of the dimensionless gate voltage $N_0$. The number 
of electrons $N$ differs appreciably from integer values only in the 
narrow \textit{mixed-valence} regions of the width
$\Delta\sim\max\lbrace\Gamma, T\rbrace/E_C$ about 
$N_0=N_0^*=\text{half-integer}$.
(b) At $\Gamma\lesssim T\ll E_C$ the conductance is small outside 
the mixed-valence regions.
}
\label{fig2}
\end{vchfigure}

We model the leads as the reservoirs of one-dimensional electrons~\cite{1D},
\beq
H_\text{leads} =  \sum_{\alpha ks}\xi^\pdag_k 
c^\dagger_{\alpha ks} c^\pdag_{\alpha ks} ,
\quad
\alpha = R,L
\label{9}
\eeq
with constant density of states $\nu$, and the tunneling between the dot 
and the leads as 
\beq
H_{\rm tunneling} 
= \sum_{\alpha k n s}t_{\!\alpha}^\pdag
c^\dagger_{\alpha k s} d^\pdag_{ns} + {\rm H.c.},
\label{10}
\eeq
where for simplicity we neglected the dependence of the tunneling amplitudes 
on $n$, so that each single-particle energy level in the dot acquires the same 
width $\Gamma_\alpha = \pi \nu t_\alpha^2$ due to the escape of electrons to 
lead $\alpha$. The width $\Gamma_\alpha$ is related to the conductance 
$G_\alpha$ of the corresponding dot-lead junction according to
$G_\alpha = (4e^2/\hbar)(\Gamma_\alpha/\delta E)$. 
The tunneling Hamiltonian description \eqref{10} is applicable when these 
conductances are small, $G_\alpha\ll e^2/h$, hence the total width 
$\Gamma = \Gamma_L+\Gamma_R$, the level spacing $\delta E$, and 
the charging energy $E_C$ form a well-defined hierarchy
\beq
\Gamma\ll\delta E\ll E_C.
\label{11}
\eeq

It is clear from \Eq{8} that \textit{half-integer} values of the 
dimensionless gate voltage are special. Indeed, consider an 
isolated dot $(\Gamma\to 0)$, so that the number of electrons 
$N$ in it is a good quantum number. The cost in electrostatic 
energy to add an extra electron to the dot then is
\[
E_{N+1} - E_N = 2E_C\bigl(N_0^\ast - N_0^\past\bigr),
\quad
N_0^\ast = N+1/2=\text{half-integer}.
\]
That is, if $N_0^\past=N_0^\ast$, the states with $N$ and $N+1$ electrons 
are degenerate. Accordingly, at low temperature $T\ll E_C$ the dependence 
of $N=\langle\hat N\rangle$ on $N_0$ should be staircase-like, see Fig.~\ref{fig2}(a),
with the width of the steps given by $\Delta\sim T/E_C$.
The charge quantization remains intact when the tunneling is now turned on, 
with the tunneling-induced width $\Gamma$ taking over from temperature at 
$T\lesssim\Gamma$. 

The quantization of charge has a profound effect on the conductance 
through the dot $G$. At very high temperature $T\gg E_C$ the 
interaction in \Eq{8} has no effect, the conductance in this limit is small, 
$G_\infty\ll e^2/h$, and is given by the classical resistance addition formula
\beq
\frac{1}{\,G_\infty}=\frac{1}{G_L}+\frac{1}{G_R}\,.
\label{12}
\eeq
Dependence on $N_0$ develops at $T\lesssim E_C$.
When the gate voltage is tuned away from the mixed-valence regions,
see Fig.~\ref{fig1}(a), an addition or a removal of an electron cost 
approximately $E_C$ in the Coulomb energy. In the course of a \textit{real} 
transition the energy must be conserved. Since for $T\ll E_C$ the 
probability to find an electron with energy $E_C$ is proportional to 
$\exp(-E_C/T)$, we expect the conductance to be exponentially 
suppressed (Coulomb blockade). On the contrary, within the 
mixed-valence regions the activation energy is small, and the 
conductance is relatively large. Thus, at $T\ll E_C$ the dependence 
$G(N_0)$ consists of a quasiperiodic sequence of narrow \textit{Coulomb 
blockade peaks} of the width $\Delta\ll 1$ separated by broad 
\textit{Coulomb blockade valleys}, as sketched in Fig.~\ref{fig2}(b).

The theory of transport in the Coulomb blockade regime was developed 
in~\cite{Shekhter}, and applied to devices with a controllable 
gate in~\cite{rate_gate}. This theory (commonly referred to as 
the \textit{orthodox theory}~\cite{orthodox}) is based on the 
rate equation formalism and is applicable at $T\gg\delta E$. 
The orthodox theory relies on the assumption that the inelastic electron 
relaxation rate within the dot is large compared to the escape rate 
$\Gamma/\hbar$~\cite{relaxation}. In other words, transitions 
between the discrete levels in the dot occur well before an electron 
escapes to the leads. Under this assumption the tunneling events 
through the two junctions can be treated independently of each other 
(this is known as the \textit{sequential tunneling} approximation; 
note that \Eq{12} relies on this approximation as well). The main 
results of the orthodox theory are summarized in Fig.~\ref{3}.

\begin{figure}[htb]
\centerline{\includegraphics[width=0.9\textwidth]{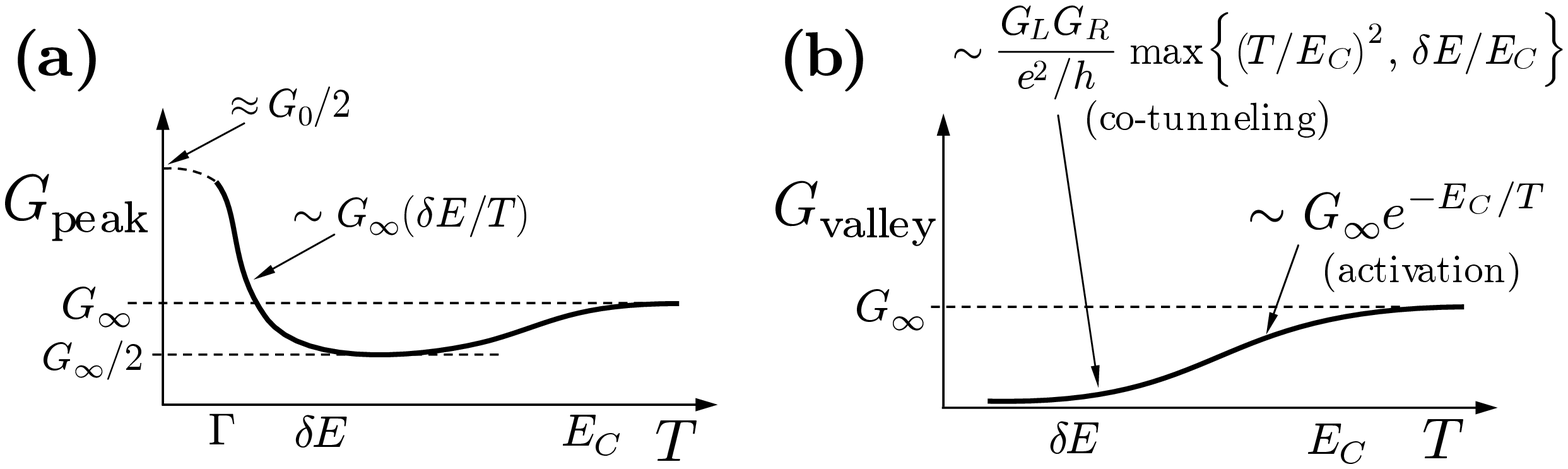}}
\caption{
(a) Temperature dependence of the height of a Coulomb blockade peak. 
(b) Conductance in the middle of a Coulomb blockade valley as function 
of temperature. 
}
\label{fig3}
\end{figure}

The orthodox theory predicts that when $T$ is lowered, the heights of the 
Coulomb blockade peaks saturate to half of their high-temperature value 
$G_\infty$. This changes if one takes into account the discreteness of the 
energy levels in the dot (the discreteness becomes important at $T\ll\delta E$). 
The rate equation approach can be used as long as $T\gg\Gamma$~\cite{rate_discrete}, 
and results in a dramatic increase of the heights of the Coulomb blockade 
peaks, see Fig.~\ref{3}(a), potentially up to $\sim e^2/h\gg G_\infty$~\cite{ABG,rate_discrete}.

\begin{figure}[htb]
\centerline{\includegraphics[width=0.85\textwidth]{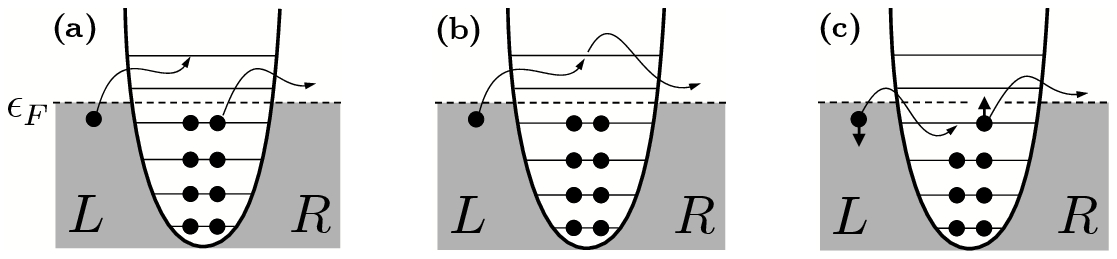}}
\caption{
A cartoon of various second-order (co-tunneling) processes~\cite{AN}.
\newline 
(a) Inelastic co-tunneling: an electron tunnels from a lead into one of 
the vacant single-particle levels in the dot, while it is an electron occupying 
some other level that tunnels out of the dot, leaving behind an electron-hole 
pair. The resulting contribution to the conductance scales with temperature 
as $T^2$. 
\newline
(b) Elastic co-tunneling: occupation numbers of single-particle energy levels
 in the dot in the initial and in the final states are exactly the same. This process
yields $T$-independent contribution to the conductance. 
\newline
(c) Elastic co-tunneling with a flip of spin, which gives rise to the Kondo effect.
}
\label{fig4}
\end{figure}

In the Coulomb blockade valleys the contribution to the conductance 
from the real transitions falls off exponentially with the decrease of 
temperature, as expected for thermally-activated transport. Eventually, 
the contributions from the higher-order (or \textit{co-tunneling}) processes 
become dominant, see Fig.~\ref{3}(b). In the co-tunneling mechanism 
states of the dot with a ``wrong'' charge participate in tunneling only as 
\textit{virtual states}, so that the conservation of energy is no longer an 
issue: the events of electron tunneling from one of the leads into the dot, 
and tunneling from the dot to the other lead occur as a single quantum 
process, see Fig.~\ref{fig4}. The existence of such higher-order 
contributions was discussed already in~\cite{Giaever}. The first 
quantitative theory of this effect, however, was developed much 
later~\cite{AN}. 

It should be noted that the co-tunneling contribution is very sensitive 
to the details of the model, and, in particular, to the dependence of the 
tunneling amplitudes in \Eq{10} on $n$. The estimate~\cite{AN} quoted 
in Fig.~\ref{fig3}(b) is appropriate for large semiconductor quantum 
dots~\cite{blockade_exp} with chaotic motion of electrons in 
it~\cite{review,ABG}. In this case the mesoscopic (valley-to-valley) 
fluctuations of the elastic co-tunneling contribution are strong, of the 
order of its average value~\cite{AG,ABG}. In fact, the elastic co-tunneling 
dominates the fluctuations of the conductance in the Coulomb blockade 
valleys at all $T\lesssim E_C$~\cite{review}.

\section{Kondo effect in a single electron transistor}

Consider now a Coulomb blockade valley with $N=\text{odd}$ electrons 
in the dot. In the ground state, the top-most occupied level is filled 
with a single electron, which may be either in a spin-up or in a spin-down 
state. In other words, the dot has a spin $S=1/2$ and its ground state 
is doubly degenerate. This singly-occupied level plays a special role 
in transport, as the elastic co-tunneling process involving this level 
may be accompanied by a flip of the transferred electron's spin with 
a simultaneous flip of the spin of the dot, see Fig.~\ref{4}(c). This is 
precisely the kind of a spin-flip process that gives rise to the Kondo 
effect in tunneling~\cite{tunneling,Appelbaum,tunneling_reviews}.
Accordingly, we expect the conductance to increase with the decrease 
of temperature as shown in Fig.~\ref{5}(a).

Let us now look more closely at the model \eqref{8}-\eqref{10}. 
Instead of dealing with the operators $c_{R,L}$ describing conduction 
electrons in the right/left leads, we can work with their linear combinations
\beq
\begin{pmatrix}
\psi\\\phi
\end{pmatrix}
=
\bigl(t_R^2 + t_L^2\bigr)^{-1/2}
\begin{pmatrix}
t_R & t_L
\\ 
- t_L & t_R 
\end{pmatrix}
\begin{pmatrix}
c_R \\ c_L
\end{pmatrix}.
\label{13} 
\eeq
With tunneling amplitudes independent of $n$, it is obvious from Eqs. \eqref{10}
and \eqref{13} that only $\psi$-electrons couple to the dot~\cite{GR-NL}. 
The corresponding part of the Hamiltonian then takes the form
\beq
H = \sum_{ks}\xi^\pdag_k 
\psi^\dagger_{ks} \psi^\pdag_{ks} 
+ \,t\sum_{kns}\bigl(\psi^\dagger_{ks} d^\pdag_{ns} + {\rm H.c.}\bigr)
+ \sum_{ns} \epsilon_n d^\dagger_{ns} d^\pdag_{ns} 
+ E_C\bigl(\hat N - N_0\bigr)^2
\label{14} 
\eeq
with the tunneling amplitude $t = \sqrt{t_R^2 + t_L^2}$. \Eq{14} is 
the simplest possible multilevel generalization of the Anderson impurity 
model~\cite{AM}. For $N_0\approx\text{odd integer}$ it can be further 
reduced to the form of the Kondo model~\Eq{3}. In this effective Kondo 
model the spin operator $\bf S$ describes the doubly-degenerate ground 
state of the dot. The exchange amplitude $J$ in \eqref{3} can be estimated as
\beq
\nu J \sim \Delta\bigl\lvert N_0\past\!\!\!-N_0^\ast\bigr\rvert^{-1} > 0, 
\label{15} 
\eeq
where $N_0^\ast$ is the half-integer number closest to $N_0$ and 
$\Delta = \Gamma/E_C$ (note that $J$ turns out to be positive). 
This estimate is easy to understand as follows. Suppose 
$N_0=\text{odd integer}$. The exchange is a result of a second-order 
process, see Fig.~\ref{fig4}(c), which involves two acts of tunneling 
(hence the factor $t^2$) and an intermediate virtual state with $N_0 \pm 1$ 
electrons in the dot. The energy of this state is $E_C$, so that 
$\nu J\sim \nu t^2\!/E_C\sim\Delta$.

\begin{figure}[htb]
\centerline{\includegraphics[width=0.85\textwidth]{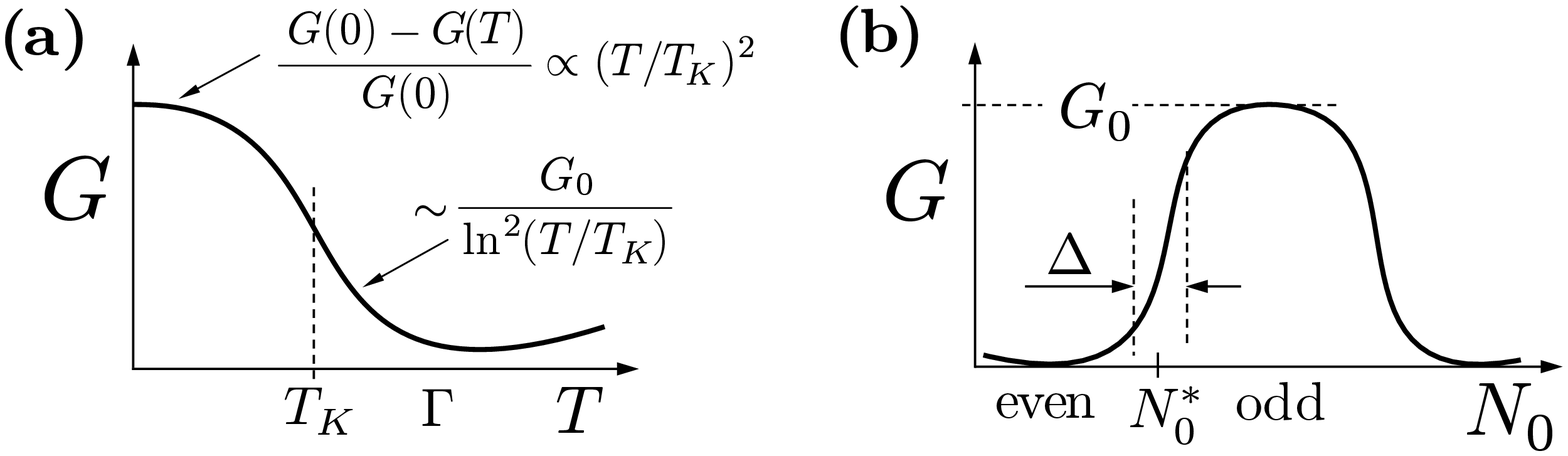}}
\caption{
(a) Temperature dependence of the conductance in the Coulomb 
blockade valley with odd number of electrons in the dot.
(b) Conductance at $T=0$ as function of the gate voltage.
}
\label{fig5}
\end{figure}

The reduction of \Eq{14} to the effective Kondo model is possible only 
when the number of electrons in the dot is odd, i.e. when the gate voltage is
outside the mixed-valence region
$\bigl\lvert N_0\past\!\!\!-N_0^\ast\bigr\rvert\lesssim\Delta$.
In addition, the temperature must be sufficiently low so that the dot can 
be safely assumed to be in the ground state.  This gives the high-energy 
cutoff for \Eq{3}
\beq
\lvert \xi_k-\epsilon_F \rvert 
< D\sim\min
\Bigl\lbrace
\delta E,\,
2E_C\bigl\lvert N_0\past\!\!\!-N_0^\ast\bigr\rvert
\Bigr\rbrace.
\label{16} 
\eeq
This condition excludes from the consideration conduction electrons 
which have enough energy to either excite an electron-hole pair in the 
dot, or to overcome the Coulomb barrier and induce a real transition 
to one of the states with ``wrong'' $N$. Since only electrons in a narrow 
strip of the width $\lvert \xi_k-\epsilon_F \rvert \sim T$ are involved in 
transport, imposing the cutoff $D\gtrsim T$ is perfectly legitimate.
Substitution of Eqs. \eqref{15} and \eqref{16} into \Eq{5} shows that
the Kondo temperature $T_K$ reaches its maximum $T_K\sim\Gamma$ 
at the border of the mixed-valence region 
$\bigl\lvert N_0\past\!\!\!-N_0^\ast\bigr\rvert\sim\Delta$,
and falls off exponentially with the increase of the distance 
$\bigl\lvert N_0\past\!\!\!-N_0^\ast\bigr\rvert$ to the charge 
degeneracy point. 

We turn now to the conductance at zero temperature. Since at any gate voltage
the ground state of the system is not degenerate (thanks to the Kondo effect!)
the conductance can be calculated with the help of the Landauer formula, 
which relates $G$ to the transmission probability through the dot.
In the conventional scattering theory approach (see, e.g.,~\cite{Newton}), 
the probability is expressed via the scattering phase shifts for the conduction 
electrons. In our toy model only $\psi$-particles scatter by the dot and 
acquire the phase shift, so that the conductance takes the form
(see~\cite{review} for the details) 
\beq
G = G_0 \frac{1}{2}\sum_s \sin^2\delta_s,
\label{17}
\eeq
where $\delta_s$ is the scattering phase shift at the Fermi level for $\psi$-particles 
with spin $s$, and 
\beq
G_0 = \frac{2e^2}{h}\left[\frac{2t_L t_R}{t_L^2 + t_R^2}\right]^2.
\label{18}
\eeq
For the Hamiltonian \eqref{14} the phase shifts can be expressed exactly 
(see, e.g.,~\cite{Langreth} or recall the Friedel sum rule) via the dot's 
occupations
\[
\delta_s = \pi N_s,
\quad
N_s = \Bigl\langle\sum_n d^\dagger_{ns}d^\pdag_{ns}\Bigr\rangle.
\] 
Since the ground state is a singlet, there is no difference between 
the spin-up and spin-down occupations, i.e. $N_s = N/2$.
\Eq{17} then yields 
\beq
G = G_0 \sin^2(\pi N).
\label{19}
\eeq
With the staircase-like dependence $N(N_0)$, see Fig.~\ref{fig2}(a),
the dependence $G(N_0)$ takes the form sketched in Fig.~\ref{5}(b). 
The conductance reaches its maximum $\max\lbrace{G\rbrace}=G_0$ 
at $N=\text{odd integer}$. The value of $G_0\leq 2e^2/h$, see \Eq{18}, 
is determined by the asymmetry of the dot-leads junctions: $G_0=2e^2/h$ 
for $t_L=t_R$~\cite{GR-NL}. Note that $G$ goes through zero at 
$N=\text{even integer}$; the appearence of these antiresonances is an 
artefact of the model \eqref{10}.  With the increase of temperature the 
conductance decreases for odd $N$, see Fig.~\ref{fig5}(a), and increases 
for even $N$. The effect of a finite $T$ is the strongest in the middle 
of the Coulomb blockade valleys with odd $N$ where $T_K$ reaches 
its minimum. 

\section{Concluding remarks}

In this short review we discussed the simplest form of the Kondo effect in a 
single electron transistor. When number of electrons in a quantum dot is odd, 
such system is essentually equivalent to the conventional $S=1/2$ magnetic 
impurity imbedded in a tunneling barrier between two massive conductors. 
The Kondo effect in this case results in the lifting of the Coulomb blockade, 
and can even yield a perfect resonant transmission at sufficiently low temperature. 
Such behavior was indeed observed in lateral GaAs quantum dots~\cite{lateral_unitary}. 

In general, the Kondo effect arises whenever a coupling to conduction electrons 
induces transitions within a degenerate ground state multiplet of an interacting 
system. Unlike the conventional magnetic impurities, nanostructures can be easily 
tuned to the degeneracy point. Perhaps the simplest such effect was predicted 
in~\cite{Zeeman} and observed in a carbon-nanotube based single-electron 
transistor in~\cite{nanotube}. Consider a quantum dot with even number of 
electrons. The ground state of such dot (which has zero spin) and one of 
the components of the excited triplet $(S=1)$ state, see Fig.~\ref{fig6}, 
become degenerate in a magnetic field with Zeeman energy $B\sim \delta E$. 
The transitions between these two states are facilitated by the elastic 
co-tunneling process analogous to that in Fig.~\ref{4}(c), and the Kondo 
effect again manifests itself in the lifting of the Colulomb blockade.

\begin{vchfigure}[htb]
\includegraphics[width=0.3\textwidth]{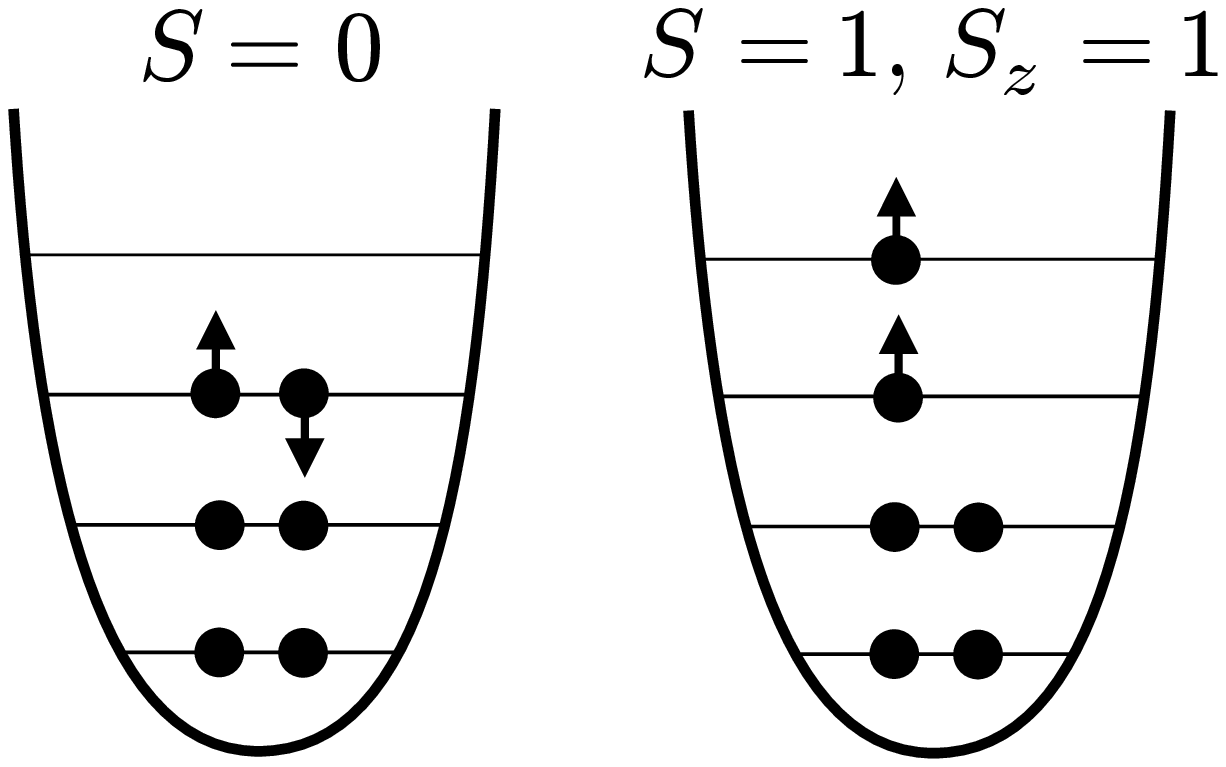}
\caption{
In a magnetic field, the ground state of a quantum dot with even $N$ 
and one of the components of the excited triplet state can become 
degenerate. 
}
\label{fig6}
\end{vchfigure}

More complicated degeneracies occur in semiconductor quantum 
dots~\cite{two-stage,vertical}. Due to the smallness of the effective mass 
in GaAs, even a very weak magnetic field applied to these systems has 
a strong orbital effect. This makes it possible to tune the device to 
the point where the singlet and the triplet states are degenerate. The theory 
of this version of the Kondo effect was developed in~\cite{ST} for the vertical 
dots and in~\cite{ST_lateral} for the lateral ones. 
When the intradot exchange interaction is sufficiently strong, the dot 
may have spin $S>1/2$ in the ground state. In this case, the Kondo 
effect occurs in two stages~\cite{real}: the conductance first raises, 
potentially up to $2e^2/h$, and then drops again as the temperature is 
further lowered. Such non-monotonic temperature dependence was 
observed in~\cite{two-stage}.

The spin degeneracy is not the only possible source of the Kondo 
effect. Any discrete index may play the part of spin, e.g. the orbital 
degeneracy in carbon nanotubes~\cite{nanotube-orbital,Borda}.
Interestingly, even the problem of finding the average charge in a 
very large quantum dot in the mixed-valence regime can be mapped 
onto the Kondo model~\cite{Matveev}, with the two almost 
degenerate charge states of the dot playing the part of the two states 
of spin-1/2. This setup turns out to be a robust realization~\cite{Matveev} 
of the symmetric (i.e. having equal exchange constants) two-channel 
$S=1/2$ Kondo model~\cite{NB}. The model results in a peculiar 
temperature dependence of the observable quantities, which at low 
temperatures follow power laws with manifestly non-Fermi-liquid 
fractional values of the exponents. Another quantum-dot based realization 
of the two-channel Kondo model has been proposed in~\cite{OGG}, and 
is now a subject of intensive research, see e.g.~\cite{2CK}.

These are just a few out of many possible extensions of the simple ideas 
discussed in this paper.

\bigskip

\begin{acknowledgement}
~\newline
This work was partially supported by the Nanoscience/Nanoengineering 
Research Program of Georgia Tech.
\end{acknowledgement}


\begin{thebibliography}{99}
\bibitem{puzzle} 
The longest-standing puzzle was, apparently, the superconductivity, 
whose explanation came half a century after its discovery. For comparison, 
it took only a few monthes to explain the fractional quantum Hall effect.
\bibitem{scales}
For example, a GaAs-based semiconductor quantum 
dot~\cite{lateral,lateral_unitary,two-stage,blockade_exp} of a linear size 
$\sim 100\text{ nm}$ contains about $10$ electrons. The charging energy 
of such dot $E_C\sim 1\text{ meV}$, while the mean single-particle level 
spacing $\delta E$ is roughly 10 times smaller. Hence, both the charging 
effects and the effects associated with the quantization of the single-particle 
energy levels can be resolved in transport experiments performed in dilution 
refrigerators with base temperatures below $50\text{ mK}$~\cite{blockade_exp}.
\bibitem{1D}
This is an excellent approximation for lateral quantum dot systems 
formed by the gate depletion of two-dimensional electron gas at the 
interface between two semiconductors~\cite{blockade_exp} (see the 
discussion in~\cite{review,ABG,real} and references therein), but 
not neseccarily for other systems of interest, see, e.g., \cite{ST}.
\bibitem{relaxation}
Note that having a finite inelastic relaxation rate requires inclusion of 
mechanisms beyond the model~\Eq{8}, e.g., electron-phonon collisions.
\end{thebibliography}
\end{document}